\documentclass{article}
\usepackage{waspaa21,amsmath,graphicx,url,times,flushend}
\usepackage{color}

\usepackage{makecell}

\newcommand*{\nDNNtf}{\mathrm{nDNN2f}}

\usepackage[pscoord]{eso-pic}
\newcommand{\placetextbox}[3]{
  \setbox0=\hbox{#3}
  \AddToShipoutPictureFG*{
    \put(\LenToUnit{#1\paperwidth},\LenToUnit{#2\paperheight}){\vtop{{\null}\makebox[0pt][c]{#3}}}%
  }%
}%

\makeatletter
\def\bstctlcite{\@ifnextchar[{\@bstctlcite}{\@bstctlcite[@auxout]}}
\def\@bstctlcite[#1]#2{\@bsphack
  \@for\@citeb:=#2\do{%
    \edef\@citeb{\expandafter\@firstofone\@citeb}%
    \if@filesw\immediate\write\csname #1\endcsname{\string\citation{\@citeb}}\fi}%
  \@esphack}
\makeatother 

\title{Controlling the Remixing of Separated Dialogue\\with a Non-Intrusive Quality Estimate}

\name{Matteo Torcoli,$^{1}$
      Jouni Paulus,$^{1,2}$
      Thorsten Kastner,$^{2}$
      Christian Uhle$^{1,2}$}
\address{$^1$ Fraunhofer Institute for Integrated Circuits IIS, 91058 Erlangen, Germany\\      
         $^2$ International Audio Laboratories Erlangen, 91058 Erlangen, Germany\sthanks{A joint institution of the Friedrich-Alexander-Universit\"{a}t Erlangen-N\"{u}rnberg (FAU) and Fraunhofer IIS, Germany.}\\matteo.torcoli@iis.fraunhofer.de\\
}

\begin{document}

\ninept
\maketitle

\begin{sloppy}

\begin{abstract}

Remixing separated audio sources trades off interferer attenuation against the amount of audible deteriorations. 
This paper proposes a non-intrusive audio quality estimation method for controlling this trade-off in a signal-adaptive manner.
The recently proposed 2f-model is adopted as the underlying quality measure, since it has been  
shown to correlate strongly with basic audio quality in source separation. 
An alternative operation mode of the measure is proposed, more appropriate when considering material with long inactive periods of the target source. 
The 2f-model requires the reference target source as an input, but this is not available in many applications. 
Deep neural networks (DNNs) are trained to estimate the 2f-model intrusively using the reference target (iDNN2f), non-intrusively using the input mix as reference (nDNN2f), and reference-free using only the separated output signal (rDNN2f). 
It is shown that iDNN2f achieves very strong correlation with the original measure on the test data (Pearson $\rho=0.99$), while performance decreases for nDNN2f ($\rho\geq0.91$) and rDNN2f ($\rho\geq0.82$).
The non-intrusive estimate nDNN2f is mapped to select item-dependent remixing gains with the aim of maximizing the interferer attenuation under a constraint on the minimum quality of the remixed output (e.g., audible but not annoying deteriorations). 
A listening test shows that this is successfully achieved even with very different selected gains (up to 23\,dB difference).
\end{abstract}


\bstctlcite{IEEEexample:BSTcontrol}  

\placetextbox{0.5}{0.08}{\fbox{\parbox{\dimexpr\textwidth-2\fboxsep-2\fboxrule\relax}{\footnotesize \centering Accepted paper. © 2021 IEEE.  Personal use of this material is permitted.  Permission from IEEE must be obtained for all other uses, in any current or future media, including reprinting/republishing this material for advertising or promotional purposes, creating new collective works, for resale or redistribution to servers or lists, or reuse of any copyrighted component of this work in other works.}}}%

\section{Introduction}
\label{sec:intro}
Many applications of audio source separation aim to attenuate the interfering background signal $b(n)$ relative to the target signal $s(n)$ given the input signal $x(n) = s(n) + b(n)$, e.g., for increasing the relative speech level in TV programs~\cite{paulus:2019}, or for aesthetic reasons in music remixing~\cite{wierstorf2017}. 
Source separation computes the estimates $\hat{s}(n)$ and $\hat{b}(n)$ and a modified output mix can be obtained as $y(n) = \hat{s}(n) + \gamma \hat{b}(n)$, with remixing gain  $0< \gamma < 1$, and  $\hat{s}(n) + \hat{b}(n) = x(n)$. 
In the following, the remixing gain $g$ is considered in dB: $g = -20\mathop{\log_{10}}(\gamma)$.
Due to the imperfection of the estimates $\hat{s}(n)$ and $\hat{b}(n)$, artifacts and distortions become audible for increasing values of $g$. 

This paper presents a method (depicted in Fig.~\ref{fig:concept}) to control the perceived sound quality of $y(n)$ by mapping the estimated quality $\hat{q}$ of $\hat{s}(n)$ to the remixing gain $g$. 
Sound quality $\hat{q}$ is quantified using a novel non-intrusive estimation of the recently-proposed 2f-model for predicting Basic Audio Quality (BAQ) for source separation \cite{ksrWaspaa19}, shown to correlate strongly with human perception \cite{torcoli:2021}. 

Also our previous work \cite{uhle:2019} proposes a method for quality control in remixing. This paper introduces main novelties such as: (i) predicting the sound quality $\hat{q}$ as an intermediate result and introducing a mapping $g = m(\hat{q})$, which can be easily adjusted according to perceptual data, (ii) employing the 2f-model, (iii) considering an improved dialogue separation method, (iv) using signals at 48\,kHz sampling rate, and (v) using deeper and waveform-based DNNs. 
Other related works apply DNNs to predict the quality of source separation, e.g., estimating Source-to-Distortion-Ratio (SDR) given the separated target signal~\cite{manilow:2017} and estimating Source-to-Artifact-Ratio (SAR) given the separated target and the mixture signal~\cite{grais:2018}, but these metrics  correlate weakly with the perception of sound quality~\cite{torcoli:2021, torcoli:2018, cano:2016, cartwright:2016}. 
Intrusive and non-intrusive estimations of speech quality  for communication applications are investigated in~\cite{Gamper:2019}. 
Non-intrusive methods for estimating various speech quality indices have been proposed in~\cite{fu2018qualitynet, catellier:2020}.
With particular focus on intelligibility of speech recorded in rooms, \cite{seetharaman:2018}~proposes the estimation of the speech transmission index without using the room impulse response.  
Computational models of intelligibility and quality have also been applied for training DNNs for source separation by using perceptually inspired loss functions~\cite{fu:2018}.  
When objectives are not differentiable, previous works have used simplified implementations~\cite{martin2018deep}, gradient approximation~\cite{zhang2018training}, DNN-based approximations~\cite{fu2020learning}, Generative Adversarial Networks (GAN)~\cite{fu2019metricgan, kawanaka:2020}, and deep reinforcement learning~\cite{koizumi:2017, koizumi:2018}.

\begin{figure}
 \centering
 \includegraphics[width=\columnwidth]{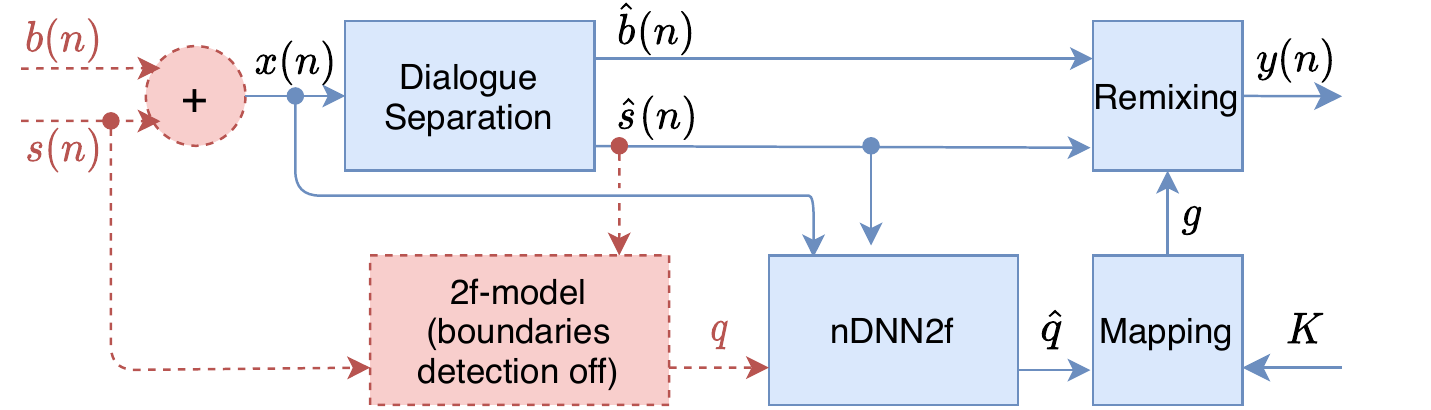}
 \caption{\label{fig:concept} A non-intrusive audio quality estimate (nDNN2f) is proposed to control the remixing trade-off between interferer reduction and final perceived quality. Dashed red lines depict training steps.}
\end{figure}

This paper is structured as follows. First, an alternative operation model of the 2f-model is proposed as underlying quality measure (Sec.~\ref{sec:2f}). 
This quality measure is intrusive, i.e., requires the reference target source, not available in the application at hand. 
Non-intrusive and reference-free DNN estimates are proposed (Sec.~\ref{sec:DNN} -- \ref{sec:data}). A mapping from a non-intrusive DNN-based estimate (nDNN2f) is proposed (Sec.~\ref{sec:mapping}) and evaluated in the remix-control application by means of a listening test (Sec.~\ref{sec:eval}).


\begin{table*}
\begin{footnotesize} 
\begin{center}
\begin{tabular}{c|c|c|c|c|c|c|c|c|c|c|c|c}
{\textbf{Layer}}
&{\textbf{In}}
&{\textbf{C1D}}
&{\textbf{C1DM}}
&{\textbf{C1DM}}
&{\textbf{C1DM}}
&{\textbf{C1DM}}
&{\textbf{C1DM}}
&{\textbf{C1DM}}
&{\textbf{Flat.}}
&{\textbf{Dense}}
&{\textbf{Dense}}
&{\textbf{Dense Out}} \tabularnewline
\hline
\makecell{\# Units \\ / Filters }
& \makecell{Waveform \\ values}
& 257
& 96
& 96
& 96
& 96
& 96
& 96
& --
& 256
& 256
& 1 \tabularnewline
\hline 
\makecell{Output \\ Shape}
&  2 or 1, 192000
& 257, 375
& 96, 188
& 96, 94
& 96, 47
& 96, 24
& 96, 12
& 96, 6
& 576
& 256
& 256
& 1 \tabularnewline
\hline 
\makecell{Filter Size \\ Stride}
& --
& \makecell{1024 \\ 512}
& \makecell{3 \\ 1}
& \makecell{3 \\ 1}
& \makecell{3 \\ 1}
& \makecell{3 \\ 1}
& \makecell{3 \\ 1}
& \makecell{3 \\ 1}
& --
& --
& --
& -- \tabularnewline
\hline 
Activation
& --
& ReLU
& ReLU
& ReLU
& ReLU
& ReLU
& ReLU
& ReLU
& --
& ReLU
& ReLU
& \makecell{hardtanh \\ $[0, 100]$}\tabularnewline
\hline 
\# Param.
& --
& 526,593  
& 74,112
& 27,744
& 27,744
& 27,744    
& 27,744
& 27,744
& 0
& 147,712
& 65,792
& 257 \tabularnewline
\end{tabular}
\end{center}
\end{footnotesize}
\vspace{-10pt}
\caption{\label{tab:arch2}The architecture of the proposed DNNs. 
\textit{C1D} stands for 1D convolutional layer. 
\textit{C1DM} is C1D followed by a max-pool layer with pooling size of 2. 
BatchNorm follows the input layer and every C1DM layer and every hidden dense layer. 
Zero-padding \textit{same} is used. 
Dense layers use 40\% dropout. 
Total number of trainable parameters is  955,880.}
\end{table*}

\section{Method}
\label{sec:method}

\subsection{Quality estimation using the adapted 2f-model}
\label{sec:2f}
The underlying computational measure for BAQ is the 2f-model as described in~\cite{ksrWaspaa19}. 
It uses two features from PEAQ \cite{thiede2000peaq} as implemented in~\cite{Kabal}. 
This original feature extraction applies a so-called boundaries detection for discarding signal segments from both probe and reference signals when the reference is effectively silent. 
If a large portion of the reference is silent, both PEAQ features and the 2f-model indicate no distortions, regardless of the content of the probe signal.
 
This issue went unnoticed during the original development of the 2f-model, since the used data always contained significant energy in the reference signal, and the discard condition was not triggered in a significant manner.
In longer real-world signals, longer pauses of the reference signals occur more often, and the 2f-model score may deviate more from the perceived quality. 

We propose to address this issue by deactivating the boundary detection in the PEAQ feature extraction.
By doing so, the audio segments with silent reference signal are not discarded, but their audio quality estimate is computed with the 2f-model, and these signals are also included in the training of the DNNs estimating the 2f-model score.
Deactivating the boundary detection does not change the results presented in~\cite{torcoli:2021} because of the non-silent nature of the reference signals considered there. 

\subsection{Predicting the 2f-model by DNNs}
\label{sec:DNN}

This work proposes using DNNs for estimating the 2f-model output in three ways: (i) intrusive, having access to the same reference signal as the original 2f-model (\textit{iDNN2f}), (ii) non-intrusive,  using the mixture signal from the dialogue separation input as the quality reference (\textit{nDNN2f}), similar to~\cite{uhle:2019}, and (iii) reference-free,  without any reference signal (\textit{rDNN2f}).
While only nDNN2f and rDNN2f can be used in the envisioned application, iDNN2f is also investigated to quantify the effect of having access to the target signal.

The network architecture is shown in Table~\ref{tab:arch2}.
It takes time-domain signals as input (i.e., it is wave-form based), in segments of 192'000 samples (corresponding to 4\,s at 48\,kHz sampling rate) from the probe signal and possibly from the reference signal.
The loss function is computed as mean squared error (MSE) between the prediction of the model and the reference 2f-model score. 
The training is run for 50 epochs using ADADELTA~\cite{zeiler2012adadelta} optimizer with initial learning rate 0.1, reduced by the factor 0.5 when the validation loss has not improved for 5 epochs. The model with the lowest validation loss is selected. 

\subsection{Traing, validation, and test data}
\label{sec:data}

The \textbf{training data} is generated with the intention of producing a range of different separation distortions, combined with a range of levels of interferer attenuation.
The training signals are created using 2\,h\,37\,min of broadcast audio with separate stems for clean speech and background (music and effects), and 1\,h\,23\,min of other clean speech samples mixed with music or noise samples as background. 
Speech and background signals are mixed at random signal-to-noise ratios (SNR) uniformly distributed in $[-10, 10]$\,dB. 
Estimates of separated speech are computed using two dialogue separation methods (\cite{paulus:2019} and an earlier version of the system described in~\cite{IBC:2021} with less training data and minor implementation differences) and adding the estimated background  scaled by $\{-\infty, -40, -14\}$\,dB, and by synthesizing artificial distortions. 
The estimated background components are obtained by subtracting the separated or simulated speech output from the input mixture. 
Original component signals with backgrounds scaled by $\{-\infty, -35, -20, 0\}$\,dB mimick artifact-free separation. 
Individual artificial distortions are applied to the original speech components with parametrization in round brackets refering to the implementation in~\cite{torcoli:2016}. These include: additive musical noise ($90\%$), low-pass filtering (3\textsuperscript{rd} order Butterworth with cut-off at 1\,kHz), clipping ($50\%$), and reducing time-frequency resolution (50\,ms, 500\,Hz). 
In addition, these artificial distortions are used with parameters randomized in ranges such that the distortions just become audible and background estimates attenuated by $45$\,dB. 
This results in total training data of 96\,h length to which 30\,min of silence is added. 
During training, each data batch of 64 examples is further duplicated with inverting the audio signal phase and copying the target 2f-model score, resulting into training batch size of 128. 

Two additional data sets are created with the same separation and augmentation.   
The \textbf{validation data} is created from 15\,min of raw data.
The \textbf{unseen test data} is created from 45\,min of broadcast material.
Even though the validation and test data sets contain items not used for the training, the separation and augmentation to obtain speech estimates was the same.  
Therefore, a fourth data set simulating an \textbf{unseen separation system} is constructed by processing the 45\,min of raw test material with the dialogue separation system presented in~\cite{IBC:2021}. 
The output component signals are combined with the background level scaled by $\{-\infty, -40, -14, 0\}$\,dB.
Additionally, the clean speech signals and the unprocessed input mixtures were used for testing. 
The authors are aware that a related separation system was used for generating a part of the training material, but verified that the outputs of the two systems sound clearly different.

\subsection{Mapping to remixing gains}
\label{sec:mapping}

Our goal is to control the trade-off between interferer attenuation and sound quality by maximizing $g$ under the constraint that $f(y(n)) \geq K$,
where $f(\cdot)$ quantifies the perceived audio quality, and $K$ is the desired quality level.
A \textit{closed-loop} solution would estimate the quality of $y(n)$, and perform a search over a range of values for $g$. 
Because of the efficiency of computing only one quality estimate per item (or segment), we propose using an \textit{open-loop} approach. 
In \cite{uhle:2019}, it was proposed to directly estimate $g$ such that a quality condition (set during training and based on an underlying objective measure) was met. 
In this paper we use an injective mapping:
$g = m(\hat{q})$,
where $\hat{q} = \nDNNtf(\hat{s}(n), x(n))$. In this way, $\hat{q}$ is an intermediate result based on an underlying objective measure and the final mapping can be adjusted using perceptual data specifically collected for this purpose. 

This assumes a monotonic relationship between $\hat{q}$  and $f(y(n))$, and that this relationship can be controlled by $g$. 
This relationship can be very complex due to psycho-acoustical factors and depending on the separation system performance. Our first approximation proposed here assumes $m(\cdot)$ to be linear.

Since designing the mapping $m(\cdot)$ requires perceptual data which we did not have in the first place, we turned to a pool of internal items produced in the context of~\cite{IBC:2021}. 
For these, the remixing gains were manually selected through a consensus between three expert engineers, aiming to maximizing the interferer attenuation while maintaining a \textit{good} final overall quality.
This data was not collected with the goal of designing  $m(\cdot)$, but our intention was to have a first rough mapping. 
Linear regression was performed, and the resulting mapping was $g = 0.71\hat{q} - 22.28$, with the range limited to $4 \leq g \leq 26$\,dB, as values outside this range were never seen in the manual mixing control. 
\begin{figure}
 \centering
 \includegraphics[width=0.8\columnwidth]{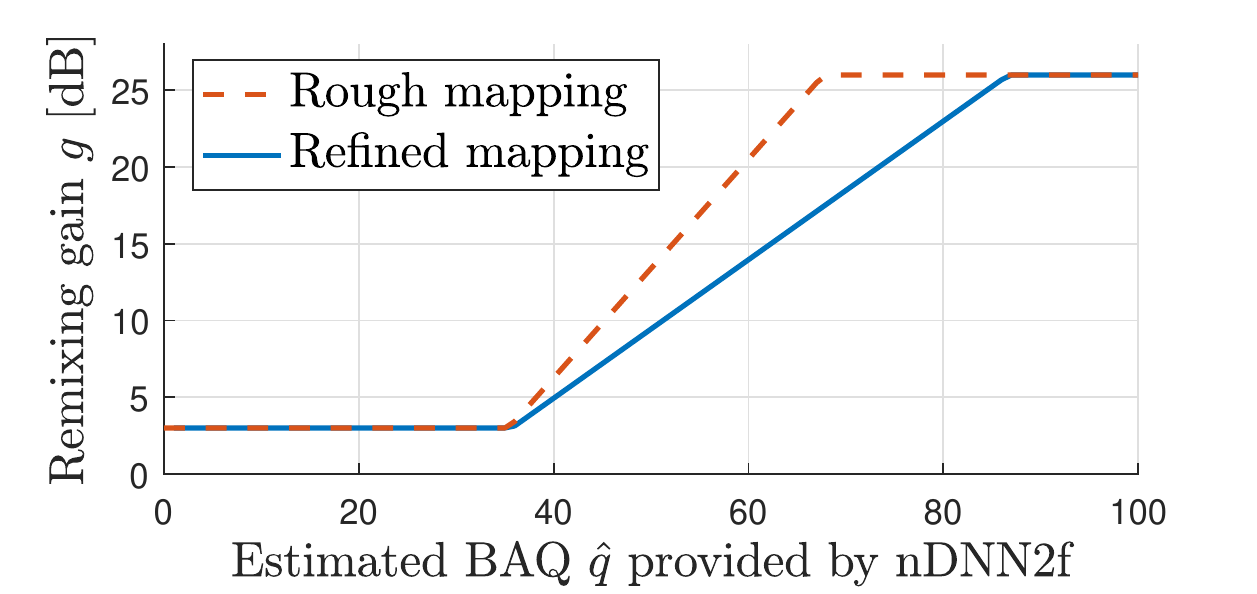}
 \caption{\label{fig:map} Proposed linear mapping from the BAQ estimate $\hat{q}$ for the separated dialogue signal $\hat{s}(n)$ to the remixing gain $g$ ($k=0$).}
\end{figure}
The mapping is illustrated by the dashed line in Fig.~\ref{fig:map}.
The possibility for further tuning this mapping was introduced by $k$, i.e., $g = m(\hat{q}) + k$.
This initial mapping was evaluated with the listening test described in Sec.~\ref{sec:LT}, and thanks to those results it was revised to be more conservative: $g  = 0.45\hat{q} - 12.67 +k$, as shown by the solid line in Fig.~\ref{fig:map}.

\section{Evaluation}
\label{sec:eval}
\subsection{Quality scores prediction performance}

Table~\ref{tab:eval_DNN} shows the performance of the proposed networks in predicting the 2f-model score on validation set, test set, and unseen separation system.
The  iDNN2f correlates with the 2f-model very strongly in all test scenarios, including the unseen system ($\rho=0.99$). 
A performance drop is observed for nDNN2f, depending on the difficulty of the task: 
it is only minimal on the validation set  ($\rho=0.98$) and it becomes more significant on the test set ($\rho=0.95$) and on the unseen system ($\rho=0.91$). 
Still, the correlation with the 2f-model is strong in all cases ($\rho\geq0.91$). 
The performance penalty for the reference-free version (rDNN2f) is bigger and the correlation with the 2f-model can decrease to $\rho=0.82$.
These performance values would decrease with active boundary detection, especially in the more challenging cases.
%

\begin{table}[t]
\begin{footnotesize}
\begin{center}
\begin{tabular}{c|c|c|c}
  & Validation set & Test set & \makecell{Test set \\+ unseen system} \\
\hline
$\rho$ & 0.99 / \textbf{0.98}  / 0.92  & 0.99 / \textbf{0.95} / 0.86  & 0.99 / \textbf{0.91} / 0.82   \\
\hline
Slope & 0.99 / \textbf{0.96}  / 0.85 & 0.98 /  \textbf{0.90} / 0.73 & 0.97 / \textbf{0.86} / 0.70  \\
\hline
MAE & 2.7 / \textbf{4.0}  / 6.6 & 2.7 / \textbf{4.9} / 7.7  & 3.1 / \textbf{8.3} / 11.3  \\
\hline
RMSE & 4.3 / \textbf{6.4}  / 11.9 & 3.9 / \textbf{7.8} / 12.8 &  4.7 / \textbf{13.35} / 17.77  \\
\end{tabular}
\end{center}
\end{footnotesize}
\vspace{-10pt}
\caption{\label{tab:eval_DNN}Pearson correlation $\rho$, linear regression slope, mean average erorr (MAE), and root mean squared error (RMSE) between DNN-based predictions and reference 2f-model. Each cell contains 3 values, referring to iDNN2f / \textbf{nDNN2f} / rDNN2f.}
\end{table}

\begin{table}[t]
\begin{footnotesize}
\begin{center}
\begin{tabular}{c|c|c|c}
 & Min $\rho$ & Max $\rho$ & Aggregated $\rho$ \\
\hline
2f-model& 0.80  & 0.92 &  0.86  \\
\hline
PEMO-Q & 0.67   & 0.84 & 0.77   \\
\hline
PESQ & 0.70 & 0.83 & 0.76   \\
\hline
POLQA (v2) & 0.64 & 0.82 & 0.74 \\
\hline
\textbf{iDNN2f} & \textbf{0.52}  & \textbf{0.75} & \textbf{0.71} \\
\hline
HAAQI& 0.59 & 0.82 & 0.69 \\
\hline
PEASS OPS & 0.54 & 0.86 & 0.67 \\
\hline
VISQOLAudio & 0.58  & 0.73  &  0.67 \\
\hline
PEAQ ODG & 0.49 & 0.75 & 0.63  \\
\hline
WEnets PESQ & 0.09  & 0.61 & 0.25 \\
\hline
\textbf{rDNN2f} & \textbf{0.09}  & \textbf{0.27} & \textbf{0.17} \\
\hline
\textbf{nDNN2f} & \textbf{0.03}  & \textbf{0.38} & \textbf{0.16} \\
\end{tabular}
\end{center}
\end{footnotesize}
\vspace{-10pt}
\caption{\label{table:wild} Minimum, maximum, and aggregated correlation $\rho$ with BAQ perceptual scores from the 6 listening tests concerning source separation. The correlation values for the proposed DNN-based estimates are shown in bold and compared with the state of the art. Full details and references for the listening tests, the state-of-the-art measures, and the correlation experiments can be found in \cite{torcoli:2021}.}
\end{table}


\begin{figure*}
 \centering
 \includegraphics[width=\textwidth]{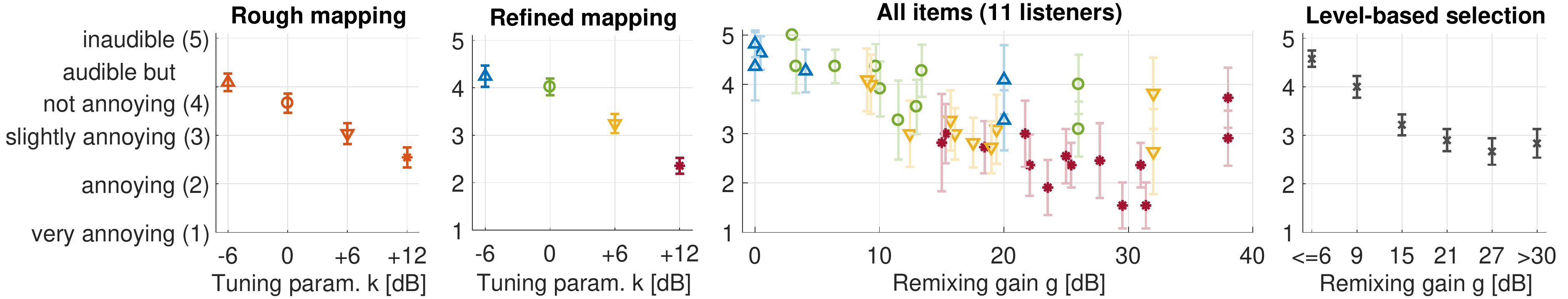}
 \caption{\label{fig:LT}Listening test results: Average perceptual scores are shown together with 95\% confidence intervals, showing that the nDNN2f quality estimate can be mapped to select non-trivial item-dependent remixing gains with desired perceived output quality. 
 }
\end{figure*}

Additionally, we considered an evaluation with perceptual scores from 6 listening tests on BAQ for source separation~\cite{torcoli:2021} to investigate the limits of the three DNNs and to test directly against perceptual reference scores. %
These listening tests involved source separation estimates of music instruments, singing voice and speech from music (the majority) and speech mixtures from various separation systems. Hence, there is a substantial departure from the training conditions.
The DNN estimates are computed over chunks of 4 seconds with 50\% overlap and the average over all segments and channels is considered.
Table \ref{table:wild} reports minimum and maximum Pearson correlation $\rho$ achieved on these listening tests together with the aggregated correlation, i.e., the average correlation in the Fisher-z-domain~\cite{torcoli:2021}. The correlation scores for the proposed networks are reported in bold along with other state-of-the-art objective measures from~\cite{torcoli:2021}.
While iDNN2f is still able to compete with the state of the art, nDNN2f and rDNN2f show basically no correlation with the perceptual reference scores. 
A similar phenomenon was observed in~\cite{torcoli:2021} for WEnets PESQ~\cite{catellier:2020}.
The authors of WEnets PESQ report Pearson correlation of 0.97 with the reference measure PESQ on their test set, but when tested on these source separation data sets, the trained network revealed much weaker correlation than PESQ. 

We can conclude that the performance of nDNN2f is satisfactory only for  the application domain and the separation systems seen during training or with little departures from them. In the following, nDNN2f is used to control the unseen separation system, for which it showed strong correlation with the reference 2f-model; iDNN2f would not be a viable solution due to the missing reference target signal, and rDNN2f showed significantly lower correlation. 

\subsection{Remixing gains prediction performance}
\label{sec:LT}

A listening test was conducted with 11 expert listeners. 
The listeners participated from home office using professional headphones. 
After an initial training session, 40 items were presented one at a time in random order. The listeners were asked to rate the absolute sound quality in terms of presence of annoying artifacts or distortions, without a reference. 
The discrete 5-point annoyance scale was employed to quantify the perceived quality $f(y(n))$. The deteriorations could be rated as 1 = very annoying, 2 = annoying, 3 = slightly annoying, \mbox{4 = audible} but not annoying, 5 = inaudible~\cite{ITUTP800:1996}. 

The 40 items were generated starting from 10 mixtures input to the unseen separation system. These were excerpts (9 to 15 sec long) from mixes professionally produced for German TV and were not present in the previous train or test sets. They consisted of female or male speech mixed with music and/or environmental noise. Average input SNR was $6.1 \pm 2.7$\,dB.
From the separated components, 4 remixing gains $g$ were generated for each of the 10 initial excerpts. 
These corresponded to $k=-6, 0, +6, +12$ as per the initial rough mapping (Fig.~\ref{fig:map}).

Results are depicted in Fig.~\ref{fig:LT}. The first subplot from the left shows the average (over items and listeners) quality level for the 4 values of $k$ together with 95\% confidence intervals (using Student's t-distribution). 
The 4 values of $k$ result in 4 distinct quality levels, with non-overlapping confidence intervals.
Considering $k=0$, the remixing gains range from 3 to 26\,dB and correspond to an average perceived quality $f(y(n))=3.7$. If better quality is required $k=-6$ could be used resulting on average in $f(y(n))=4.1$. If more quality degradation is acceptable, $k=+6$ or $+12$ can be selected for more interferer reduction, with average $f(y(n))=3.0$ and $2.5$. 

The average quality for $k=0$ ($f(y(n))=3.7 \pm 0.77$) can be further improved by a refined mapping.
The refined mapping was obtained by considering only the items rated with average quality $f(y(n))\geq 3.5$. A linear regression was performed between the remixing gains for these items and the corresponding nDNN2f scores, obtaining the mapping introduced in Sec.\,\ref{sec:mapping} and Fig.~\ref{fig:map}.

For each of the 10 initial excerpts, the remixing gains in the listening test are re-clustered based on the closest gain obtained with the refined mapping and $k=-6, 0, +6, +12$.
The corresponding quality scores are depicted in the second and third subplot of Fig.~\ref{fig:LT}, where different colors and symbols encode the belonging to the categories $k=-6, 0, +6, +12$.
The second subplot shows the average scores over items and listeners, while the third subplot shows the average quality scores for the 40 individual items, averaged over listeners only.
Also with this refined mapping, the gains corresponding to $k=0$ range from 3 to 26\,dB, but the average quality is higher and the standard deviation smaller ($f(y(n))=4.0 \pm 0.58$) than with the rough mapping. Moreover, no item is scored on average below 3 for $k=0$, as can be observed in the third subplot (green circles).

The fourth subplot shows the scores from the listening test aggregated by the absolute value of the remixing gains in steps of 6\,dB (e.g., $9$ on the x-axis refers to the range $(6, 12]$ dB). It can be observed that only smaller gains (${\leq12}$\,dB) guarantee good perceptual quality (average $f(y(n))\geq4$), while the confidence intervals become bigger and overlap for $g > 12$, leaving uncertainty regarding the final item-dependent quality for this upper range.
A similar average quality to $k=0$ (refined mapping) could be selected by considering only gains in the range $(6, 12]$\,dB. However, with the nDNN2f-based selection, 4 gains (out of 10) are $>$ 12\,dB. 
Moreover, the controlled gains obtained with $k=0$ work as quality anchors, i.e., they can be tuned with 6\,dB steps, obtaining consistent changes in the perceived quality, which is observed only to a limited extent in the level-based selection.
Hence, nDNN2f can be mapped to select non-trivial item-dependent remixing gains with desired output quality, even if the selected gains are very different in absolute terms (up to 23 dB difference).

As a final note, the same linear regression done for the refined mapping was performed by considering PEASS APS~\cite{emiya:2011}, which was the core measure used in~\cite{uhle:2019}. For $k=0$, this gave lower average quality and higher standard deviation ($f(y(n))=3.7 \pm 0.79$), using the original APS, i.e., with the full reference signals  and without the DNN-based estimation. This shows that the 2f-model is a better underlaying measure for this purpose than APS.

\section{Conclusion}
DNNs were investigated for predicting the quality of blindly separated dialogue in a non-intrusive way. As underlying quality measure, an alternative operation mode of the 2f-model was proposed. 
This quality prediction was mapped to select item-dependent remixing gains so to obtain a final modified mix where the relative dialogue boost is maximized under a quality control constrain, e.g., that the distortions introduced by the source separation may be just audible but not annoying.
The mapping between the quality estimate provided by the DNN and the remixing gain can be easily tuned, obtaining consistent changes in the perceived quality of the final remixed signals, as verified by the conducted listening test.

\bibliographystyle{IEEEtran}
\bibliography{Bibliography}
%
%
%
%
%
%
%
%
%

\end{sloppy}
\end{document}